\documentstyle[aps,manuscript]{revtex}
\begin{document}
\baselineskip 1.5pc
\begin{center}
{\Large \bf Effects of the $U_A(1)$ Anomaly on $\eta \to 2 \gamma$ Decay}
\vskip 10mm
M.~Takizawa\footnote{E-mail address: takizawa@ins.u-tokyo.ac.jp}

{\it Institute for Nuclear Study, University of Tokyo,}

{\it Tanashi, Tokyo 188, Japan}

M.~Oka\footnote{E-mail address: oka@th.phys.titech.ac.jp}

{\it Department of Physics, Tokyo Institute of Technology,}

{\it Megro, Tokyo 152, Japan}
\end{center}
\begin{abstract}
\baselineskip 1.5pc
    We study the $\eta \to 2 \gamma$ decay using an extended three-flavor
Nambu-Jona-Lasinio model that includes the 't~Hooft instanton induced
interaction.  The $\eta$ meson mass and the $\eta \to 2 \gamma$ decay width
are reproduced simultaneously with a rather strong instanton induced
interaction.
The calculated $\eta$ decay constant is $f_\eta = 2.23 f_\pi$ and it
suggests that the $\eta$ meson is no longer the Goldstone boson.
\end{abstract}
\section{Introduction}
    The structure of the $\eta$ meson gives us information about the
mechanisms of spontaneous breaking of chiral symmetry, the pattern of
the explicit breaking of chiral symmetry and the dynamics of the $U_A(1)$
anomaly.  The $\eta$ meson is the eighth member of the low-lying nonet
pseudoscalar mesons and is considered as a Goldstone boson associated with the
spontaneous breaking of chiral symmetry in the QCD vacuum.
\par
    The physics of the $\eta$ and $\eta'$ mesons have been extensively studied
in the $1/N_C$ expansion approach \cite{tHooft74}. In the $N_C \to \infty$
limit, the $U_A(1)$ anomaly is turned off and then the $\eta$ meson becomes to
degenerate with the pion and the $\eta'$ meson becomes pure $\bar ss$ state
with
$m_{\eta'}^2 (N_C \to \infty) = 2 m_K^2 - m_\pi^2 \simeq (687 \,{\rm MeV})^2$
\cite{Vene79}.  So the $U_A(1)$ anomaly pushes up $m_{\eta}$ by about 400MeV
and $m_{\eta'}$ by about 300MeV.  Usually the $\eta'$ meson is not considered
as a Goldstone boson since it is considered as mostly flavor singlet meson and
in this channel there exists the $U_A(1)$ anomaly which explicitly breaks the
chiral symmetry.
On the other hand, the $\eta$ meson is considered as a Goldstone boson because
its mass is close to the naive estimation of the pseudoscalar meson mass in
$\lambda^8$ channel, i.e., $m_{\eta_8}^2 = \frac{4}{3} m_K^2 - \frac{1}{3}
m_\pi^2 \simeq (567 \,{\rm MeV})^2$.
{}From the $1/N_C$ expansion point of view,
$\eta$-meson mass is largely affected by the finite $N_C$ correction, i.e.,
the $U_A(1)$ anomaly, then it is natural to ask how much $\eta$ meson
loses the Goldstone boson nature.
\par
    In order to answer to the above question, it may be important to study
the $\eta \to 2 \gamma$ decay because of relation to the Adler-Bell-Jackiw
(ABJ) triangle anomaly \cite{ABJ69} through the partial conservation of
axial-vector current (PCAC) hypothesis.
\par
    One of the useful and widely used frameworks for studying the phenomena
related to the axial-vector anomaly and the spontaneous chiral symmetry
breaking is the chiral effective meson lagrangian given by Wess and Zumino
\cite{WZ71} and developed by Witten \cite{Wit83}.
The $\eta,\eta' \to 2 \gamma$ decays have been studied using the
Wess-Zumino-Witten (WZW) lagrangian with the corrections at one-loop order
in the chiral perturbation and it has been shown that the two-photon decay
widths can be explained with the $\eta$-$\eta'$ mixing angle
$\theta \simeq -20^\circ$ \cite{DHL85}.
{}From the chiral perturbation \cite{GL84} point of view, the WZW term is
derived in the chiral limit and is of order $p^4$.  As discussed in
\cite{GL85}, to reliably calculate $SU(3)$ breaking effects of the
$\eta, \eta' \to 2 \gamma$ decays, the low-energy expansion to order $p^6$
has to be carried out.  However in \cite{DHL85} full analysis of order $p^6$
has not been performed.  Furthermore, because of the $U_A(1)$ anomaly, the
singlet channel decay amplitude $\eta_0 \to 2 \gamma$ derived using PCAC +
ABJ anomaly should be modified so as to become the renormalisation group
invariant \cite{SV92}.
\par
    The purpose of this paper is to study the $\eta \to 2 \gamma$ decay in the
framework of the generalized Nambu-Jona-Lasinio (NJL) model \cite{NJL61} as
a chiral effective quark lagrangian of the low-energy QCD.  The generalized
three-flavor NJL model which involves the $U_L(3) \times U_R(3)$ symmetric
four-quark interaction and the six-quark flavor-determinant interaction
\cite{KKM71} incorporating effects of the $U_A(1)$ anomaly is used widely
in recent years to study such topics as the quark condensates in vacuum, the
spectrum of low-lying mesons, the flavor-mixing properties of the low-energy
hadrons, etc. \cite{KH88,BJM88,KKT88,RA88,KLVW90}.  In this approach the
effects of the explicit breaking of the chiral symmetry by the current quark
mass term and the $U_A(1)$ anomaly on the $\eta \to 2 \gamma$ decay amplitude
can be calculated consistently with those on the $\eta$-meson mass, $\eta$
decay constant and mixing angle within the model applicability.
\section{Extended Nambu-Jona-Lasinio model}
    We work with the NJL model lagrangian density extended to three-flavor
case:
\begin{eqnarray}
{\cal L} & = & {\cal L}_0 + {\cal L}_4 + {\cal L}_6 , \label{njl1} \\
{\cal L}_0 & = & \bar \psi \,\left( i \partial_\mu \gamma^\mu - \hat m
\right) \, \psi \, ,
\label{njl2} \\
{\cal L}_4 & = & {G_S \over 2} \sum_{a=0}^8 \, \left[\, \left(
\bar \psi \lambda^a \psi \right)^2 + \left( \bar \psi \lambda^a i \gamma_5
\psi \right)^2 \, \right] \, ,
\label{njl3} \\
{\cal L}_6 & = & G_D \left\{ \, {\rm det} \left[ \bar \psi_i (1 - \gamma_5)
\psi_j \right] + {\rm det}  \left[ \bar \psi_i (1 + \gamma_5) \psi_j
\right] \, \right\} \, .
\label{njl4}
\end{eqnarray}
Here the quark field $\psi$ is a column vector in color, flavor and Dirac
spaces and $\lambda^a$ is the $U(3)$ generator in flavor space.  The free
Dirac lagrangian ${\cal L}_0$ incorporates the current quark mass matrix
$\hat m = {\rm diag}(m_u, m_d, m_s)$ which breaks the chiral
$U_L(3) \times U_R(3)$ invariance explicitly. ${\cal L}_4$ is a QCD
motivated four-fermion interaction, which is chiral $U_L(3) \times U_R(3)$
invariant.  The 't Hooft determinant ${\cal L}_6$ represents the $U_A(1)$
anomaly.  It is a $3 \times 3$ determinant with respect to flavor with
$i,j = {\rm u,d,s}$.
\par
    Quark condensates and constituent quark masses are self-consistently
determined by the gap equations
\begin{eqnarray}
M_u & = & m_u - 2 G_S \langle \bar u u \rangle - 2 G_D \langle \bar d d
\rangle \langle \bar s s \rangle \, , \nonumber \\
M_d & = & m_d - 2 G_S \langle \bar d d \rangle - 2 G_D \langle \bar s s
\rangle \langle \bar u u \rangle \, , \nonumber \\
M_s & = & m_s - 2 G_S \langle \bar s s \rangle - 2 G_D \langle \bar u u
\rangle \langle \bar d d \rangle \, , \label{gap}
\end{eqnarray}
with
\begin{equation}
\langle \bar a a \rangle = - {\rm Tr}^{(c,D)} \, \left[ i S_F^a (x = 0) \right]
=  - \int^{\Lambda} \frac{d^4p}{(2 \pi)^4} {\rm Tr}^{(c,D)} \left[
\frac{i}{p_{\mu} \gamma^{\mu} - M_a + i \varepsilon} \right] \, . \label{cond}
\end{equation}
Here the covariant cutoff $\Lambda$ is introduced to regularize the
divergent integral and ${\rm Tr}^{(c,D)}$
means trace in color and Dirac spaces.
\par
    The pseudoscalar channel quark-antiquark scattering amplitudes
$\langle p_3 , \bar p_4 ; {\rm out} | p_1 , \bar p_2 ; {\rm in} \rangle =
(2 \pi)^4 \delta^4(p_3 + p_4 - p_1 - p_2) {\cal T}_{q \bar q}$
are then calculated in the ladder approximation. We assume the isospin
symmetry too.  In the $\eta$ and $\eta'$ channel, the explicit expression is
\begin{equation}
{\cal T}_{q \bar q} = -
\left(
\begin{array}{c}
\bar u(p_3) \lambda^8 i \gamma_5 v(p_4) \\
\bar u(p_3) \lambda^0 i \gamma_5 v(p_4)
\end{array}
\right)^T \,
\left(
\begin{array}{cc}
A(q^2) & B(q^2) \\
B(q^2) & C(q^2) \\
\end{array}
\right) \,
\left(
\begin{array}{c}
\bar v(p_2) \lambda^8 i \gamma_5 u(p_1) \\
\bar v(p_2) \lambda^0 i \gamma_5 u(p_1)
\end{array}
\right) \, , \label{qas1}
\end{equation}
with
\begin{eqnarray}
A(q^2) & = & \frac{2}{{\rm det}{\bf D}(q^2)}
\left\{ 2 ( G_0 G_8 - G_m G_m ) I^0 (q^2) - G_8 \right\} \, , \label{qas2} \\
B(q^2) & = & \frac{2}{{\rm det}{\bf D}(q^2)}
\left\{- 2 ( G_0 G_8 - G_m G_m ) I^m (q^2) - G_m \right\} \, , \label{qas3} \\
C(q^2) & = & \frac{2}{{\rm det}{\bf D}(q^2)}
\left\{ 2 ( G_0 G_8 - G_m G_m ) I^8 (q^2) - G_0 \right\} \, , \label{qas4}
\end{eqnarray}
and
$G_0 = \frac{1}{2} G_S - \frac{1}{3} ( 2 \langle \bar uu \rangle +
\langle \bar ss \rangle ) G_D$,
$G_8 = \frac{1}{2} G_S - \frac{1}{6} ( \langle \bar ss \rangle - 4
\langle \bar uu \rangle ) G_D$,
$G_m = - \frac{1}{3 \sqrt{2}} ( \langle \bar ss \rangle -
\langle \bar uu \rangle ) G_D$.
The quark-antiquark bubble integrals are
\begin{eqnarray}
I^0(q^2) & = & i \int^{\Lambda} \frac{d^4p}{(2 \pi)^4} {\rm Tr}^{(c,f,D)}
\left[ S_F(p) \lambda^0 i \gamma_5 S_F(p+q) \lambda^0 i \gamma_5 \right]
\, , \label{int1} \\
I^8(q^2) & = & i \int^{\Lambda} \frac{d^4p}{(2 \pi)^4} {\rm Tr}^{(c,f,D)}
\left[ S_F(p) \lambda^8 i \gamma_5 S_F(p+q) \lambda^8 i \gamma_5 \right]
\, , \label{int2} \\
I^m(q^2) & = & i \int^{\Lambda} \frac{d^4p}{(2 \pi)^4} {\rm Tr}^{(c,f,D)}
\left[ S_F(p) \lambda^0 i \gamma_5 S_F(p+q) \lambda^8 i \gamma_5 \right]
\, , \label{int3}
\end{eqnarray}
with $q = p_1 + p_2$.  The $2 \times 2$ matrix ${\bf D}$ is
\begin{equation}
{\bf D}(q^2) =
\left(
\begin{array}{cc}
D_{11}(q^2) & D_{12}(q^2) \\
D_{21}(q^2) & D_{22}(q^2)
\end{array}
\right) \, , \label{mat}
\end{equation}
with
\begin{eqnarray}
D_{11}(q^2) & = & 2 G_8 I^8(q^2) + 2 G_m I^m(q^2) - 1 \, , \label{mat11}\\
D_{12}(q^2) & = & 2 G_8 I^m(q^2) + 2 G_m I^0(q^2) \label{mat12} \\
D_{21}(q^2) & = & 2 G_0 I^m(q^2) + 2 G_m I^8(q^2) \label{mat21} \\
D_{22}(q^2) & = & 2 G_0 I^0(q^2) + 2 G_m I^m(q^2) - 1 \, . \label{mat22}
\end{eqnarray}
{}From the pole position of the scattering amplitude Eq. (\ref{qas1}), the
$\eta$-meson mass $m_{\eta}$ is determined.
\par
    The scattering amplitude Eq. (\ref{qas1}) can be diagonalized by rotation
in the flavor space
\begin{eqnarray}
{\cal T}_{q \bar q} & = & -
\left(
\begin{array}{c}
\bar u(p_3) \lambda^8 i \gamma_5 v(p_4) \\
\bar u(p_3) \lambda^0 i \gamma_5 v(p_4)
\end{array}
\right)^T   {\bf T}_{\theta}^{-1} {\bf T}_{\theta}
\left(
\begin{array}{cc}
A(q^2) & B(q^2) \\
B(q^2) & C(q^2) \\
\end{array}
\right) {\bf T}^{-1}_{\theta}  {\bf T}_{\theta}
\left(
\begin{array}{c}
\bar v(p_2) \lambda^8 i \gamma_5 u(p_1) \\
\bar v(p_2) \lambda^0 i \gamma_5 u(p_1)
\end{array}
\right) \, , \label{qasm1} \\
& = & -
\left(
\begin{array}{c}
\bar u(p_3) \lambda^{\eta} i \gamma_5 v(p_4) \\
\bar u(p_3) \lambda^{\eta'} i \gamma_5 v(p_4)
\end{array}
\right)^T \,
\left(
\begin{array}{cc}
D^{\eta}(q^2) & 0 \\
0 & D^{\eta'}(q^2)
\end{array}
\right) \,
\left(
\begin{array}{c}
\bar v(p_2) \lambda^{\eta} i \gamma_5 u(p_1) \\
\bar v(p_2) \lambda^{\eta'} i \gamma_5 u(p_1)
\end{array}
\right) \, , \label{qasm2}
\end{eqnarray}
with $\lambda^{\eta} \equiv \cos \theta  \lambda^8 - \sin \theta \lambda^0$,
$\lambda^{\eta'} \equiv \sin \theta  \lambda^8 + \cos \theta \lambda^0$ and
$$
{\bf T}_{\theta} = \left(
\begin{array}{cc}
\cos \theta & -\sin \theta \\
\sin \theta & \cos \theta
\end{array}
\right) \, .
$$
The rotation angle $\theta$ is determined by
\begin{equation}
\tan 2 \theta = \frac{2 B(q^2)}{C(q^2) - A(q^2)} \, . \label{angle}
\end{equation}
So $\theta$ depends on $q^2$.  At $q^2 = m_{\eta}^2$,  $\theta$ represents the
mixing angle of the $\lambda^8$ and $\lambda^0$ components in the $\eta$-meson
state.  In the usual effective pseudoscalar meson lagrangian approaches, the
$\eta$ and $\eta'$ mesons are analyzed using the $q^2$-independent
$\eta$-$\eta'$ mixing angle.  Because of the $q^2$-dependence, $\theta$ cannot
be interpreted as the $\eta$-$\eta'$ mixing angle.  The origin of the
$q^2$-dependence is that the $\eta$ and $\eta'$ meson have the internal
structures.
\par
    The effective $\eta$-quark coupling constant $g_{\eta}$ is determined
by the residue of the scattering amplitude at the $\eta$ pole, i.e.,
$g_{\eta}^2 = \lim_{q^2 \to m_{\eta}^2} (q^2 - m_{\eta}^2) D_{\eta}(q^2)$,
and the $\eta$ decay constant $f_\eta$ is determined by calculating the
quark-antiquark one-loop graph,
\begin{equation}
f_{\eta} = \frac{g_{\eta}}{m_{\eta}^2} \int^{\Lambda} \frac{d^4p}{(2 \pi)^4}
{\rm Tr}^{(c,f,D)} \bigl[q^{\mu} \gamma_{\mu} \gamma_5 \frac{\lambda^{\eta}}{2}
S_F(p) i \gamma_5 \lambda^{\eta} S_F(p-q)\bigr] \left\vert_{q^2 = m_{\eta}^2}
\right. \, . \label{edc}
\end{equation}
\par
    One can easily show that in the $U_A(1)$ limit, i.e., $G_D = 0$ and
$m_{u,d} \not= m_s$, the $\eta$ meson becomes the ideal mixing state composed
of u and d-quarks, namely, $m_{\eta} = m_{\pi}$, $g_{\eta} = g_{\pi}$,
$f_{\eta} = f_{\pi}$ and $\tan \theta = - \sqrt{2}$.
\section{$\eta \rightarrow 2 \gamma$ decay amplitudes}
    Let us now turn to the calculations of the $\pi^0, \eta \to 2 \gamma$
decay widths.  The  $\pi^0, \eta \to 2 \gamma$ decay amplitudes are given
by
\begin{equation}
\langle \gamma (k_1) \gamma (k_2) \vert M (q) \rangle = i (2 \pi)^4
\delta^4 (k_1 + k_2 -q) \varepsilon_{\mu \nu \rho \sigma}
\epsilon^{\mu}_1 \epsilon^{\nu}_2 k_1^{\rho} k_2^{\sigma}
\widetilde{\cal T}_{M \to 2 \gamma}(q^2) \, , \label{ampl}
\end{equation}
where $\epsilon_1$ and $\epsilon_2$ are the polarization vectors of the
photon.  By calculating the pseudoscalar-vector-vector type quark triangle
diagrams, we get the following results.
\begin{eqnarray}
\widetilde{\cal T}_{\pi^0 \to 2 \gamma} & = & \frac{\alpha}{\pi} g_{\pi}
 F(u,\pi^0) \, , \label{amplpi}  \\
\widetilde{\cal T}_{\eta \to 2 \gamma} & = & \frac{\alpha}{\pi} g_{\eta}
\frac{1}{3\sqrt{3}} \left[ \cos \theta \left\{ 5 F(u,\eta) - 2 F(s,\eta)
\right\} - \sin \theta \sqrt{2} \left\{ 5 F(u,\eta) + F(s,\eta) \right\}
\right] \, .  \label{ampleta}
\end{eqnarray}
Here $\alpha$ is the fine structure constant of QED and $F(a,M)$ ($a=u,s$ and
$M=\pi^0,\eta$) is defined as
\begin{equation}
F(a,M) = \int_0^1 dx \int_0^1 dy \,
         \frac{2 (1-x) M_a}{M_a^2 - m_M^2 x (1-x) (1-y)} \, . \label{fam}
\end{equation}
Then the $M \to 2 \gamma$ decay width $\Gamma(M \to 2 \gamma)$ is given by
$\Gamma(M \to 2 \gamma) = \vert \widetilde{\cal T}_{M \to 2 \gamma} \vert^2
m_M^3 /(64 \pi)$.
\par
    In the chiral limit, the pion mass vanishes and $F(u,\pi^0)$ becomes
$1/M_u$.  In this limit, the Goldberger-Treiman (GT) relation at the quark
level, $M_u = g_{\pi} f_{\pi}$, holds in the NJL model and this leads to
$\widetilde{\cal T}_{\pi^0 \to 2 \gamma} = \alpha/(\pi f_{\pi})$ which is
same as the tree-level results in the WZW lagrangian approach.  It should be
mentioned that we have to integrate out the triangle diagrams without
introducing a cutoff $\Lambda$ in order to get the above result  though the
cutoff is introduced in Eqs. (\ref{cond},\ref{int1}-\ref{int3},\ref{edc})
in the NJL model.   If we introduce a cutoff $\Lambda$ to the
loop-integral of the triangle diagrams, the decay amplitude becomes too small
and we lose the success of the model independent prediction for the
$\Gamma(\pi^0 \to 2 \gamma)$.  In the $U(3)_L \times U(3)_R$ version of the
NJL model, the WZW term has been derived using the bosonization method with
the heat-kernel expansion \cite{ER86,W89}.  In thier approach, $O(1/\Lambda)$
term has been neglected and it is same as to take the $\Lambda \to \infty$
limit.
\section{Numerical Results}
    The recent experimental results of the $\pi^0, \eta \to 2 \gamma$ decay
 widths are
$\Gamma(\pi^0 \to 2\gamma) = 7.7 \pm 0.6 \, {\rm eV}$ and
$\Gamma(\eta \to 2\gamma) = 0.510 \pm 0.026 \, {\rm keV}$ \cite{PDG} and the
reduced amplitudes are
\begin{eqnarray}
\left\vert \widetilde{\cal T}_{\pi^0 \to 2 \gamma} \right\vert & = &
(2.5 \pm 0.1) \times 10^{-11} \, [{\rm eV}]^{-1} \, , \label{exppidw} \\
\left\vert \widetilde{\cal T}_{\eta \to 2 \gamma} \right\vert & = &
(2.5 \pm 0.06) \times 10^{-11} \, [{\rm eV}]^{-1} \, . \label{expetadw}
\end{eqnarray}
{}From Eq. (\ref{amplpi}) and Eq. (\ref{ampleta}), we get
$\widetilde{\cal T}_{\eta \to 2 \gamma} = (5/3)
\widetilde{\cal T}_{\pi^0 \to 2 \gamma}$ in the $U_A(1)$ limit.  Therefore
in order to reproduce the experimental value of
$\widetilde{\cal T}_{\eta \to 2 \gamma}$, the effect of the $U_A(1)$ anomaly
should reduce $\widetilde{\cal T}_{\eta \to 2 \gamma}$ by a factor 3/5.
\par
    In our theoretical calculations, the parameters of the NJL model are the
current quark masses $m_u = m_d$, $m_s$, the four-quark coupling constant
$G_S$, the coupling constant of the 't~Hooft instanton induced interaction
$G_D$ and the covariant cutoff $\Lambda$.  First, we take $G_D$ as a free
parameter and study the $\eta$-meson properties as functions of $G_D$.
For the light quark masses, $m_u = m_d = 8.0$ MeV is taken to reproduce
$M_u = M_d \simeq 330$ MeV which is the value usually used in the
nonrelativistic quark model.  Other parameters $m_s$, $G_D$ and $\Lambda$ are
determined so as to reproduce the isospin averaged observed masses,
$m_\pi = 138.04$ MeV, $m_K = 495.7$ MeV and the pion decay constant $f_\pi$.
\par
    The calculated constituent u,d-quark mass is $M_{u,d} = 324.9$ MeV which
is independent of $G_D$.  On the other hand the calculated constituent
s-quark mass weakly decreases from $M_s = 556.3$ MeV to $M_s = 503.3$ MeV when
$G_D$ is changed from $G_D = 0$ to $G_D = G_D^\eta$ where the observed
$m_\eta$ is reproduced.  The fitted result of the current s-quark mass $m_s$
is almost independent of $G_D$ and $m_s = 192.95$ MeV at $G_D = G_D^\eta$.
The ratio of the current s-quark mass to the current u,d-quark mass is
$m_s/m_u = 24.1$, which agrees well with $m_s/\hat{m} = 25 \pm 2.5$
 ($\hat{m} = \frac{1}{2} (m_u + m_d)$) derived from ChPT \cite{GL82}.
The kaon decay constant $f_K$ is the prediction and is almost independent of
$G_D$.  We have obtained $f_K = 96.6$ MeV at $G_D = G_D^\eta$ which is about
15\% smaller than the observed value.  We consider this is the typical
predictive power of the NJL model in the strangeness sector.
\par
    We next discuss the $\pi^0 \to 2 \gamma$ decay.  The calculated result
is $\widetilde{\cal T}_{\pi^0 \to 2 \gamma} = 2.50 \times 10^{-11} 1/{\rm eV}$
which agrees well with the observed value given in Eq. (\ref{exppidw}).
The current algebra result is
$\widetilde{\cal T}_{\pi^0 \to 2 \gamma} = \alpha/(\pi f_\pi) = 2.514 \times
10^{-11} 1/{\rm eV}$, so the soft pion limit is a good approximation for
$\pi^0 \to 2 \gamma$ decay.  There are two effects of the symmetry breaking
on $\widetilde{\cal T}_{\pi^0 \to 2 \gamma}$.  One is the deviation from
the G-T relation and another is the matrix element of the triangle diagram
$F(u,\pi^0)$.  Our numerical results are $g_\pi = 3.44$, $M_u/f_\pi = 3.52$
and $F(u,\pi^0) M_u = 1.015$, therefore the deviations from the soft pion
limit are very small both in the G-T relation and the matrix element of the
triangle diagram.
\par
    Let us now turn to the discussion of the $\eta$-meson properties. The
calculated results of the $\eta$-meson mass $m_\eta$ and the mixing angle
$\theta$ are shown in Fig. 1, the $\eta$ decay constant $f_\eta$ is given in
Fig. 2 and the $\eta \to 2 \gamma$ decay amplitude
$\widetilde{\cal T}_{\eta \to 2 \gamma}$ is given in Fig. 3 as functions of
the non-dimensionalized coupling constant of 't~Hooft instanton induced
interaction $G_D^{eff} \equiv - G_D (\Lambda/2\pi)^4 \Lambda N_C^2$.
\par
    As can be seen from Fig. 1, in order to reproduce the observed
$\eta$-meson mass, rather strong instanton induced interaction is necessary.
For example, at $G_D = G_D^\eta$, $G_D \langle \bar s s \rangle/G_s = 1.58$,
it means that the contribution of ${\cal L}_6$ to the dynamical mass of the
u,d-quarks is about 60\% bigger than that of  ${\cal L}_4$.  In the previous
study of the $\eta$ and $\eta'$ mesons in the extended NJL model
\cite{KH88,BJM88,KKT88,RA88,KLVW90}, the strength of the instanton induced
interaction has been determined so as to reproduce the observed $\eta'$
mass though the $\eta'$ state has the unphysical decay mode of the
$\eta' \to \bar uu, \bar dd$.  The strength determined from $m_{\eta'}$,
$G_D^{\eta'}$ is much smaller than $G_D^\eta$, about 1/10 to 1/5 of $G_D^\eta$.
One of the shortcomings of the NJL model is the lack of the confinement
mechanism.  It is expected that the confinement gives rise to the attractive
force between quark and antiquark in the $\eta'$ meson to prevent the $\eta'$
meson from decaying to the quark and antiquark pair.
\par
    One of the important results is that the calculated $\eta$ decay constant
is very different from the pion decay constant,
$f_\eta = 206\,{\rm MeV} = 2.23 f_\pi$ at $G_D = G_D^\eta$. This suggests
that the $\eta$ meson loses the Goldstone boson nature very much.
To see whether the $\eta$ meson is the Goldstone boson, the G-T relation is
another important information.  For the $\eta$ meson, the naive G-T relation
at the quark level is
$2 g_\eta f_\eta = | \frac{2}{\sqrt{3}} \cos \theta
- 2 \sqrt{\frac{2}{3}} \sin \theta | M_u + | - \frac{2}{\sqrt{3}}
\cos \theta - \sqrt{\frac{2}{3}} \sin \theta | M_s$
and at
$G_D = G_D^\eta$, our numerical results are
$2 g_\eta f_\eta = 3.202$ GeV and
$| \frac{2}{\sqrt{3}} \cos \theta - 2 \sqrt{\frac{2}{3}} \sin \theta | M_u +
 | - \frac{2}{\sqrt{3}} \cos \theta - \sqrt{\frac{2}{3}} \sin \theta | M_s =
 0.892$ GeV.
Therefore the G-T relation dose not hold at all.
Since the NJL model is known as the model that describes the Goldstone boson
properties reasonably well, it is natural to ask whether the present model
is applicable to the $\eta$ meson.  For the $\eta'$ meson, we expect that the
confinement plays an important role since the $\eta'$-meson pole of the
scattering amplitude Eq. (\ref{qas1}) appears above the $\bar uu$ and
$\bar dd$-thresold.  On the other hand, the $\eta$ meson is the tight bound
state, so we expect that the NJL model can describe the essential feature of
the $\eta$ meson.  In order to confirm it, we have studied the constituent
u,d-quark mass dependence of the $\eta$-meson properties.  By changing
$m_{u,d}$ from 7.5 MeV to 8.5 MeV, we have changed $M_{u,d}$ from about
300 MeV to 360 MeV and other parameters of the model have been chosen so as
to reproduce the experimental values of $m_\pi$, $m_K$, $m_\eta$ and $f_\pi$.
The changes of the calculated $\eta$ decay constant and the mixing angle have
been within 2\%.
\par
    We are now in the position to discuss the $\eta \to 2 \gamma$ decay
amplitude.  Our result is
$\widetilde{\cal T}_{\eta \to 2 \gamma} = 2.73 \times 10^{-11}$ 1/eV at
$G_D = G_D^\eta$, which is about 10\% larger than the experimental value.
Therefore the present model reproduces the $\eta$-meson mass and the
$\eta \to 2 \gamma$ decay width simultaneously.
As for the effects of the symmetry breaking on
$\widetilde{\cal T}_{\eta \to 2 \gamma}$, our results are
$F(u,\eta) M_u = 1.41$ and $F(u,\eta)/F(s,\eta) = 1.96$.
Bernard et.al. \cite{BBHMR93} calculated the $\eta \to 2 \gamma$ decay width
using a similar model.  They used a rather weak instanton induced
interaction and their result of
$\Gamma(\eta \to 2 \gamma)$ is about 50\% bigger than the experimental value.
It is understandable from our analysis.
\par
    Our calculated result of the mixing angle is $\theta = 15.1^\circ$ which
should be compared with $\theta \simeq -20^\circ$ \cite{GK87}.  There are two
major differences between our model calculations and the usual analysis.  One
is that, in the usual analysis, the energy independent mixing is assumed.
Another point is that the effects of the $U_A(1)$ anomaly on the $\eta$-meson
properties are not taken into account in the usual analysis.
\section{Concluding Remarks}
    Using an extended three-flavor Nambu-Jona-Lasinio model that includes the
't~Hooft instanton induced interaction, we have studied the
$\pi^0, \eta \to 2 \gamma$ decays as well as the properties of the pion,
the kaon and the $\eta$ meson.  The $\eta$-meson mass and the
$\eta \to 2 \gamma$ decay width
have been reproduced simultaneously with a rather strong instanton induced
interaction.  The calculated $\eta$ decay constant is about twice of the pion
decay constant.  So it is rather hard to consider the $\eta$ meson as the
Goldstone boson.  Because of the above novel picture of the $\eta$ meson,
the situation of the mixing angle is also different from the usual analysis.
\par
    In order to confirm the novel picture of the $\eta$ meson we have obtained
here, certainly further studies of the $\eta$-meson properties are necessary.
One thing is to study other $\eta$-meson decay processes such as
$\eta \to \pi^0 2 \gamma$, $\eta \to 3 \pi$ in the present framework and such
calculations are now in progress.  Since the properties of the $\eta$ meson
and those of the $\eta'$ meson are closely related, one has to construct the
low-energy effective model of QCD which can apply to the $\eta'$ meson.
Such an attempt is left as future study.

\pagebreak
\vskip 3mm
\centerline{\bf Figure Captions}
\vskip 5mm
\newcommand{\namelistlabel}[1]{\mbox{#1}\hfil}
\newenvironment{namelist}[1]{%
\begin{list}{}
      {\let\makelabel\namelistlabel
       \settowidth{\labelwidth}{#1}
       \setlength{\leftmargin}{1.1\labelwidth}}
}{%
\end{list}}
\begin{namelist}{fig. 1xx}
\item[{\bf Fig. 1}]Dependence of the calculated $\eta$-meson mass $m_\eta$
(solid line) and the mixing angle $\theta$ (dashed line) on the
non-dimensionalized coupling constant of the 't~Hooft instanton induced
interaction $G_D^{eff}$.  The observed $\eta$-meson mass $m_\eta = 547.45$
MeV is reproduced at $G_D^{eff} = 1.41$.
\vskip 3mm
\item[{\bf Fig. 2}]Dependence of the calculated $\eta$ decay constant
$f_\eta$ on the non-dimensionalized coupling constant of the 't~Hooft
instanton induced interaction $G_D^{eff}$.
\vskip 3mm
\item[{\bf Fig. 3}]Dependence of the $\eta \to 2 \gamma$ decay amplitude
$\widetilde{\cal T}_{\eta \to 2 \gamma}$ on the
non-dimensionalized coupling constant of the 't~Hooft instanton induced
interaction $G_D^{eff}$.
\end{namelist}

\begin{thebibliography}{99}
\bibitem{tHooft74}G. 't Hooft, Nucl. Phys. {\bf B72} (1974) 461; \hfil\break
For a review; G.A. Christos, Phys. Rep. {\bf 116} (1984) 251.
\bibitem{Vene79}G. Veneziano, Nucl. Phys. {\bf B159} (1979) 213.
\bibitem{ABJ69}S. Adler, Phys. Rev. {\bf 177} (1969) 2426; \hfil\break
J. Bell and R. Jackiw, Nuovo Cimento {\bf 60A} (1969) 47.
\bibitem{WZ71}J. Wess and B. Zumino, Phys. Lett. {\bf B37} (1971) 95.
\bibitem{Wit83}E. Witten, Nucl. Phys. {\bf B223} (1983) 422.
\bibitem{DHL85}J.F. Donoghue, B.R. Holstein and Y.-C.R. Lin, Phys. Rev. Lett.
{\bf 55} (1985) 2766; Erratum {\bf 61} (1988) 1527; \hfil\break
J. Bijnens, A. Bramon and F. Cornet, Phys. Rev. Lett. {\bf 61} (1988) 1453;
\hfil\break
J.F. Donoghue and D. Wyler, Nucl. Phys. {\bf B316} (1989) 289.
\bibitem{GL84}J. Gasser and H. Leutwyler, Ann. of Phys. {\bf 158} (1984) 142.
\bibitem{GL85}J. Gasser and H. Leutwyler, Nucl. Phys. {\bf B250} (1985) 465.
\bibitem{SV92}G.M. Shore and G. Veneziano, Nucl. Phys. {\bf B381} (1992) 3.
\bibitem{NJL61}Y. Nambu and G. Jona-Lasinio, Phys. Rev. {\bf 122} (1961) 345;
{\bf 124} (1961) 246.
\bibitem{KKM71}M. Kobayashi, H. Kondo and T. Maskawa, Prog. Theor. Phys.
{\bf 45} (1971) 1955; \hfil\break
G. 't Hooft, Phys. Rev. {\bf D14} (1976) 3432.
\bibitem{KH88}T. Kunihiro and T. Hatsuda, Phys, Lett. {\bf B206} (1988) 385;
\hfil\break T. Hatsuda and T. Kunihiro, Z. Phys. {\bf C51} (1991) 49;
Phys. Rep. {\bf 247} (1994) 221.
\bibitem{BJM88}V. Bernard, R.L. Jaffe and U.-G. Meissner, Nucl. Phys.
{\bf B308} (1988) 753.
\bibitem{KKT88}Y. Kohyama, K. Kubodera and M. Takizawa, Phys. Lett. {\bf B208}
(1988) 165; \hfil\break
M. Takizawa, K. Tsushima, Y. Kohyama and K. Kubodera, Prog. Theor. Phys.
{\bf 82} (1989) 481; Nucl. Phys. {\bf A507} (1990) 611.
\bibitem{RA88}H. Reinhardt and R. Alkofer, Phys. Lett. {\bf B207} (1988) 482;
\hfil\break R. Alkofer and H. Reinhardt, Z. Phys. {\bf C45} (1989) 275.
\bibitem{KLVW90}S. Klimt, M. Lutz, U. Vogl and W. Weise, Nucl. Phys.
{\bf A516} (1990) 429; \hfil\break
U. Vogl, M. Lutz, S. Klimt and W. Weise, Prog. Part. Nucl. Phys. {\bf 27}
(1991) 195.
\bibitem{ER86}D. Ebert and H. Reinhardt, Nucl. Phys. {\bf B271} (1986) 188.
\bibitem{W89}M. Wakamatsu, Ann. of Phys. {\bf 193} (1989) 287.
\bibitem{PDG}Particle Data Group, Phys. Rev. {\bf D50} (1994) 1173.
\bibitem{GL82}J. Gasser and H. Leutwyler, Phys. Rep. {\bf 87} (1982) 77.
\bibitem{BBHMR93}V. Bernard, A.H. Blin, B. Hiller, U.-G. Meissner and
M.C. Ruivo, Phys. Lett. {\bf B305} (1993) 163.
\bibitem{GK87}F.J. Gilman and R. Kauffman, Phys. Rev. {\bf D36} (1987) 2761.
\end{thebibliography}
\end{document}